\documentclass[journal=amrcda, article=account]{achemso}
\usepackage[bookmarks=false]{hyperref} 
\usepackage{chemformula}

\author{Hengrui Zhang}
\affiliation{Department of Mechanical Engineering, Northwestern University, Evanston, IL 60208, United States}

\author{Alexandru B.~Georgescu}
\affiliation{Department of Materials Science and Engineering, Northwestern University, Evanston, IL 60208, United States}
\alsoaffiliation{Department of Chemistry, Indiana University, Bloomington, IN 47405, United States}

\author{Suraj Yerramilli}
\affiliation{Department of Industrial Engineering and Management Sciences, Northwestern University, Evanston, IL 60208, United States}

\author{Christopher Karpovich}
\affiliation{Department of Materials Science and Engineering, Massachusetts Institute of Technology, Cambridge, MA 02139, United States}

\author{Daniel W.~Apley}
\affiliation{Department of Industrial Engineering and Management Sciences, Northwestern University, Evanston, IL 60208, United States}

\author{Elsa A.~Olivetti} 
\affiliation{Department of Materials Science and Engineering, Massachusetts Institute of Technology, Cambridge, MA 02139, United States}

\author{James M.~Rondinelli}
\affiliation{Department of Materials Science and Engineering, Northwestern University, Evanston, IL 60208, United States}
\email{jrondinelli@northwestern.edu}

\author{Wei Chen}
\affiliation{Department of Mechanical Engineering, Northwestern University, Evanston, IL 60208, United States}
\email{weichen@northwestern.edu}

\title{Emerging microelectronic materials by design: Navigating combinatorial design space with scarce and dispersed data}

\begin{document}

\begin{tocentry}
    \includegraphics[width=\textwidth]{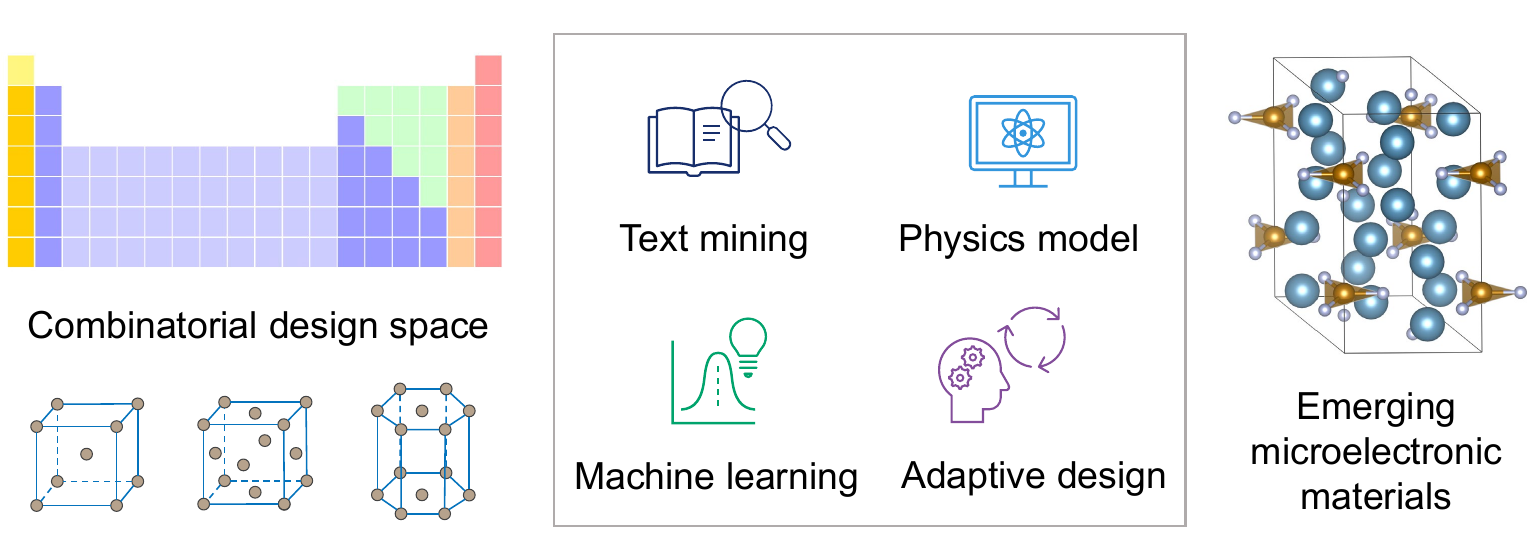}
\end{tocentry}

\begin{abstract}
The increasing demands of sustainable energy, electronics, and biomedical applications call for next-generation functional materials with unprecedented properties. Of particular interest are emerging materials that display exceptional physical properties, making them promising candidates in energy-efficient microelectronic devices. As the conventional Edisonian approach becomes significantly outpaced by growing societal needs, emerging computational modeling and machine learning methods are employed for the rational design of materials. However, the complex physical mechanisms, cost of first-principles calculations, and the dispersity and scarcity of data pose challenges to both physics-based and data-driven materials modeling. Moreover, the combinatorial composition--structure design space is high-dimensional and often disjoint, making design optimization nontrivial.

In this Account, we review a team effort toward establishing a framework that integrates data-driven and physics-based methods to address these challenges and accelerate materials design. We begin by presenting our integrated materials design framework and its three components in a general context. (1) Using text mining and natural language processing techniques, our framework first extracts and organizes relevant information dispersed in the literature. (2) From this initial database of relevant materials, data-driven models can be trained and subsequently employed to perform virtual screening of the unknown materials space. This virtual screening process can identify promising materials families for further investigation, thus narrowing down the candidate space. (3) Within the identified materials families, a Bayesian optimization-based adaptive discovery workflow is applied to search for materials with optimal properties. To extend the capability of Bayesian optimization, which was previously restricted to small data and numerical variables, we develop a family of uncertainty-aware machine learning methods for mixed numerical and categorical variables.

We then provide an example of applying this materials design framework to metal--insulator transition (MIT) materials, a specific type of emerging materials with practical importance in next-generation memory technologies. We identify multiple new materials that may display this property in the lacunar spinel and Ruddlesden-Popper perovskite families, and propose pathways for their synthesis. The classifiers used to identify new possible MIT materials also identified previously unknown features that may be used for a predictive theory for this class of materials. For example, we have identified descriptors derived from ionicity and atom sizes of materials as key to MIT behavior.

Finally, we identify some outstanding challenges in data-driven materials design, such as materials data quality issues, property--performance mismatch, and validation and deployment. We seek to raise awareness of these overlooked issues hindering materials design, thus stimulating efforts toward developing methods to mitigate the gaps.
\end{abstract}

\section{Introduction}
The growing energy demand for computing that poses increasing challenges to climate change, as well as the desire for novel computing devices beyond Moore's law, urge the need for next-generation energy-efficient computing technologies. Beyond continually improving current microelectronic materials, a technological breakthrough requires emerging materials with unprecedented properties. For example, metal--insulator transition (MIT) materials \cite{mit-revmp-98}, which can switch between conductive and insulating through a rapid electronic phase transition, are promising alternatives to the current (structural) phase change memory (PCM) materials in microelectronic devices. However, the current pace of conventional trial-and-error materials discovery makes it impossible to fulfill these societal needs \cite{tb-joule-18}.
With recent developments in computational modeling, artificial intelligence, and design automation, the materials science community seeks ``materials by design'' \cite{arroyave-anrev-19}, i.e., systematically and rationally optimizing materials that meet the demand without serendipity. Mainly enabled by physics-based predictive modeling and data-driven design methods \cite{rgb-accounts-22}, this methodology has facilitated the development of various materials to address pressing needs in advanced technologies \cite{npj-newfront-19}.

Despite the growing number of successes, the design of materials faces significant challenges. The design space for potential materials formed by chemical composition and structure usually has high dimensionality. Even among defect-free crystalline materials, the various crystal structures and configurations of different atoms within these structures lead to a combinatorial explosion: thousands, or even millions, of configurations, could be possible at the atomic scale, forming an enormous design space. Physics-based computational models, particularly first-principles models based on density functional theory (DFT), are expensive and time-consuming.
This makes it prohibitive to evaluate all candidates in the design space. Finally, these materials are often scarce: for example, when we began the project, the number of known materials with a thermally-driven MIT was under 70, and spread in several materials families such as perovskite, spinel, and rutile.

Complement to the physics-based methods, data-driven design utilizes existing data for materials property prediction and optimization. Machine learning (ML) can surrogate costly simulations to predict materials’ properties and performance accurately yet inexpensively \cite{mldesign-arcbe-22}, thus facilitating ``virtual screening'' that rapidly locates promising candidates out of an enormous design space. On the other hand, adaptive design optimization techniques such as Bayesian optimization (BO) \cite{bo-pieee-16} allow for efficiently finding the global optimal design starting from a small amount of data. It has been applied to the design optimization of a wide spectrum of materials and processes \cite{bo-cheme-22}.

One challenge hindering the application of data-driven methods is the disjoint design space. Mathematically, data-driven design is viewed as the optimization of several design variables. Numerical variables are well handled by common ML models and are easy to optimize. However, materials design usually involves mixed categorical (e.g., element type) and numerical (e.g., fraction of element) variables. Not only does the mixed-variable complexity pose difficulty to ML models, but it also makes the design space disjoint, thus challenging optimization techniques such as BO. This calls for ML models that can handle mixed variables and support BO.

Another challenge is associated with data acquisition. Data for specific functional materials of interest could be accumulated over decades of studies. However, the previous findings are reported in literature across publishers, topics, and even fields. This dispersity makes collecting initial data a bottleneck. Text mining and natural language processing (NLP) techniques have been applied to extract knowledge of materials from literature \cite{olivetti-nlp-20} and assist in synthesis planning \cite{mehr-science-20}. These successes show an extraordinary potential of mining data from the literature to promote materials design, which has yet to be systematically exploited.

In this Account, we summarize the methods we have developed towards addressing the challenges and propose a systematic framework (Figure \ref{fig:frame}) for designing materials in the combinatorial space using scarce and dispersed data. We demonstrate our approach in the design of MIT materials. As a general approach to combinatorial materials design, the framework applies to other materials systems as well. The sections are organized as follows: Section \ref{sec:frame} presents an overview of our proposed framework and the MIT materials design problem. Sections \ref{sec:textmining}--\ref{sec:adaptive} describe the three major components of the framework. Section \ref{sec:outlook} discusses remaining challenges and outlooks in accelerating materials design in the combinatorial space.

\section{Framework} \label{sec:frame}
\begin{figure}[ht]
    \centering
    \includegraphics[width=\textwidth]{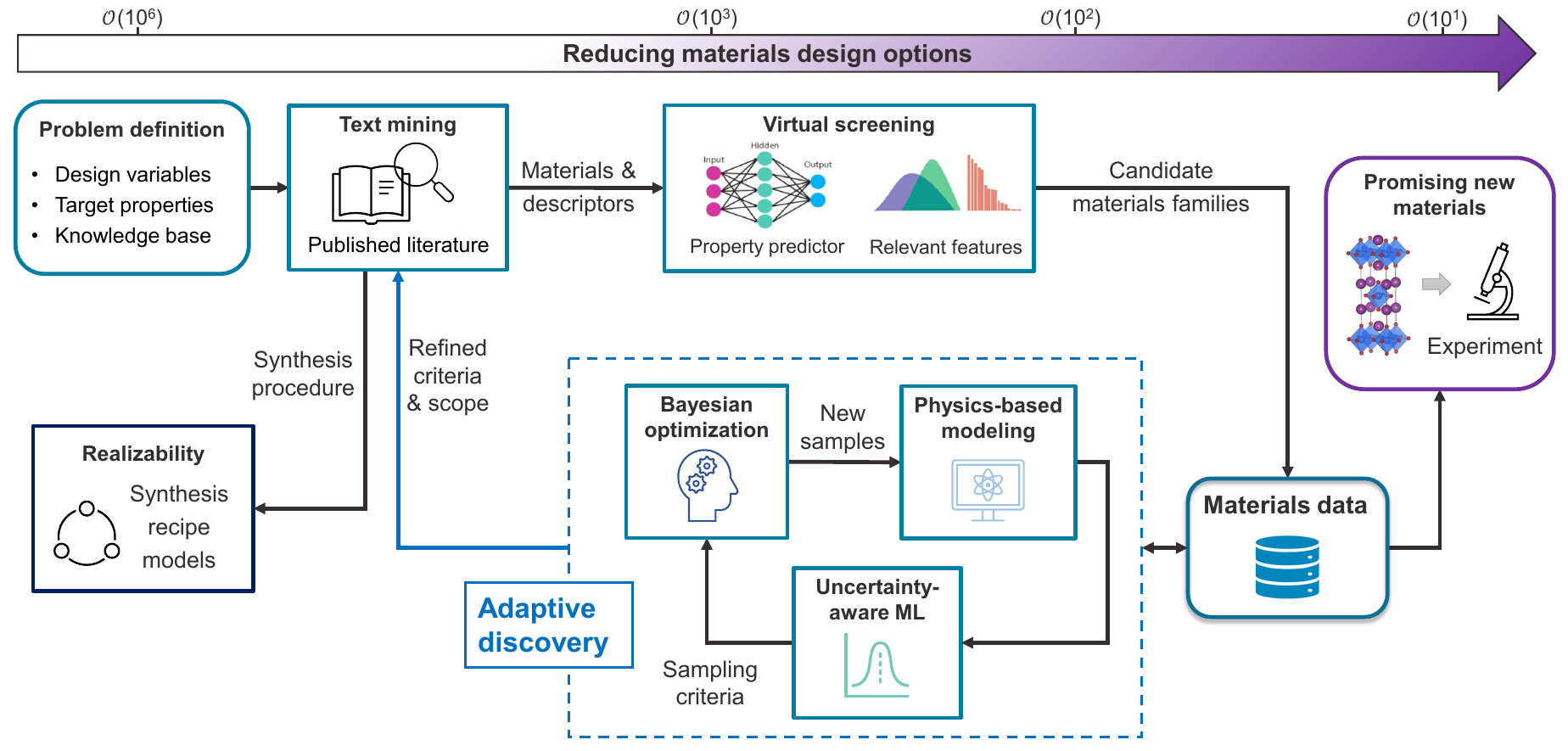}
    \caption{An integrated computational materials design framework.
    (1) The problem definition specifies design variables, target properties, and a knowledge base from published literature. Text mining techniques help to extract relevant data from the literature. (2) Using the data, ML models are trained and employed to virtually screen the vast materials databases to identify candidate materials families. (3) BO and physics-based computational models are combined to search for an optimal material within the candidate families.}
    \label{fig:frame}
\end{figure}

Figure \ref{fig:frame} shows the overall workflow of our proposed materials design framework. At the problem formulation stage, the design variables (input) and target properties (output) are determined based on the intended application and design requirements, and prior knowledge of the scope and physical mechanisms may be available. These guide the collection of relevant literature, from which text mining can extract information including investigated materials, key descriptors, material properties, and synthesis procedures.
ML models are trained on the extracted data. The models can predict target properties of unseen materials, thus enabling virtual screening of vast databases. Through virtual screening, promising candidate material families are identified, decomposing the intractable large design space into smaller sub-spaces. The design options could be reduced to thousands, or hundreds, from a design space of over $10^6$. 

Afterward, adaptive design optimization combining Bayesian optimization (BO) and physics-based simulations is performed within each identified material family. Adaptive design methods, such as BO, could guide computational simulation toward the optimal solution (sets), thus significantly reducing the number of samples to be simulated and the computational cost. The design options are thus reduced from $10^3$ to $10^1$.
Based on knowledge learned from this adaptive discovery process, new design criteria and scope could be defined, thus initiating a new design process.
The technologies that make up this framework are introduced in the following sections.

\begin{figure}[ht]
    \centering
    \includegraphics[width=0.8\textwidth]{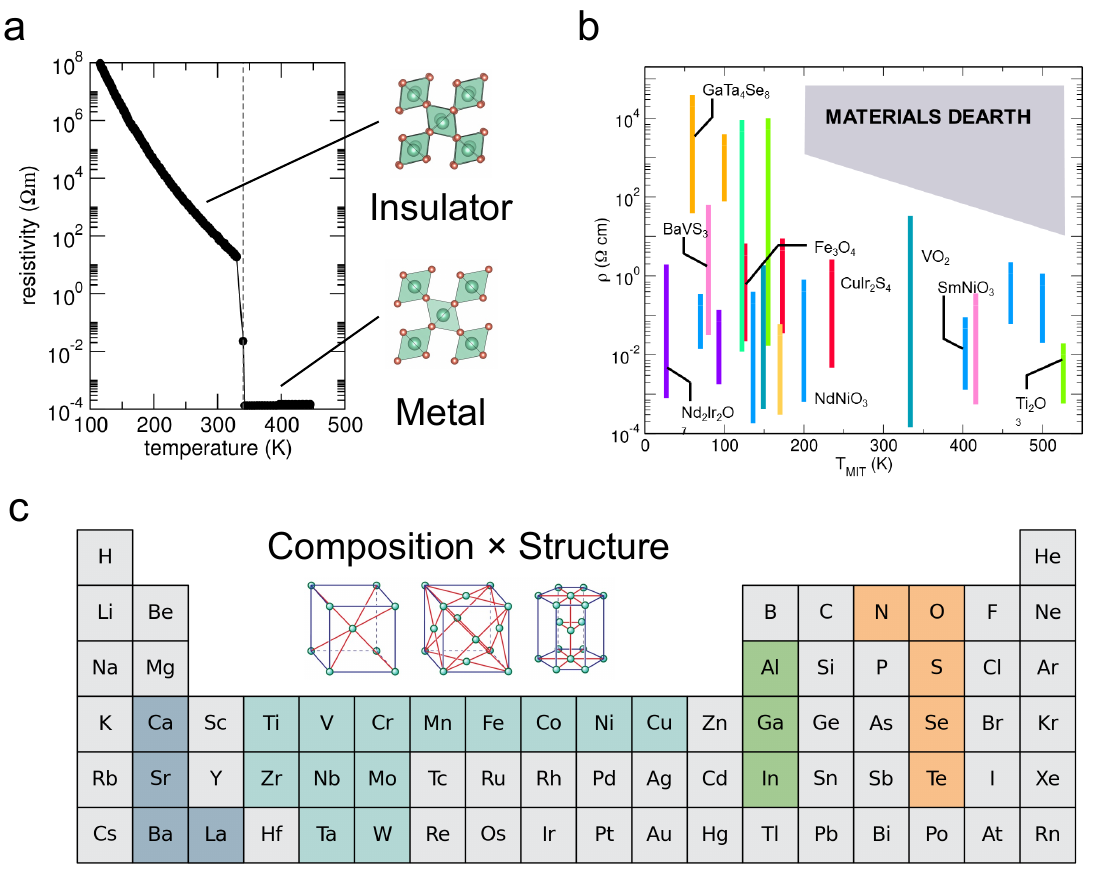}
    \caption{(a) An illustration of MIT material: upon temperature change across a critical value, the material undergoes a structural distortion, which leads to a drastic change in electrical resistivity. (b) Some previously discovered MIT materials, with their transition temperature $T_{\mathrm{MIT}}$ and resistivity change range shown. A material dearth is present for the desired large resistivity change at high temperatures. (c) The expansive design space formed by the combination of composition and structure.}
    \label{fig:challenges}
\end{figure}

As a demonstration of the proposed materials design framework, we apply it to the design of MIT materials \cite{mit-revmp-98}, which are promising for energy-efficient microelectronic devices, including neuromorphic computing devices \cite{parker-am-22}, Mott field effect transistors \cite{shukla-nc-15}, radio-frequency antenna, and cross-point selector memories (Figure \ref{fig:challenges}).

MIT materials can undergo rapid changes in electrical resistivity (by orders of magnitude) upon external stimuli, such as a temperature change or electrical field. This property enables information encoding and decoding at a higher rate and lower energy consumption. While some MIT materials, such as \ch{VO2}, have been discovered and applied in devices \cite{Mahanta_2018,Bohaichuk_2019} none of the known MIT materials achieve a desired reversible resistivity change $\sim 10^5$ near room temperature, which is required to outperform silicon-based and PCM devices \cite{Zhang_2015}.
On the other hand, the discovery of new MIT materials following the conventional approach cannot fulfill the demand timely, and the large atomic structure--composition design space (Figure \ref{fig:challenges}c) of prospective MIT materials is mostly unexplored.
Our proposed framework provides a systematic and efficient way of searching for high-performance synthesizable MIT materials in this large design space to promote the development of next-generation microelectronic devices.

\section{Knowledge Extraction from Literature} \label{sec:textmining}
As the available data relevant to MITs are sparse and dispersed, we start by gathering the data from the literature using text mining, which enables subsequent data-driven studies.
\subsection{Text Mining Pipeline}
\begin{figure}[ht]
    \centering
    \includegraphics[width=0.8\linewidth]{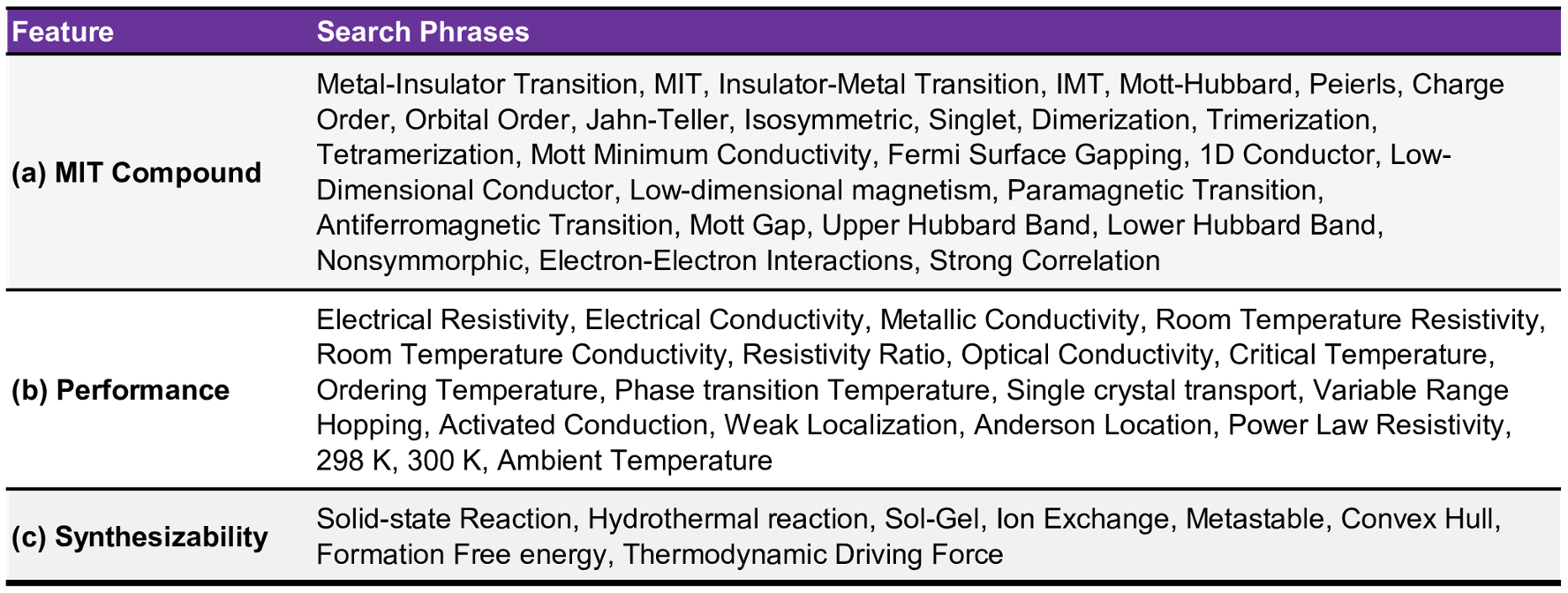}
    \caption{A list of keywords related to MIT materials, their performances, and synthesis.}
    \label{fig:mit-keywords}
\end{figure}

Prior knowledge related to the materials of interest may be scattered in an enormous text corpus of millions of scientific publications. This initial large corpus could be narrowed down by keyword-directed searching. To do so, we first define a list of keywords relevant to compounds, performance, and synthesis. For MIT materials, keywords could include physical phenomena (correlated electrons, Jahn-Teller, etc), performance metrics (resistivity), synthesis methods (hydrothermal reaction, molecular beam epitaxy), and known candidate families (perovskite, spinel). Figure \ref{fig:mit-keywords} provides a more comprehensive list of keywords we utilized for text mining focused on MIT materials. Following the workflow established in our team's previous works \cite{nlp-scidata-17,elsa-acscs-19}, we employed a variety of text parsing and pattern matching algorithms to find papers whose title, abstract, main text, or figure captions contain the keywords, thus curating a specialized text corpus. Searching in an initial corpus of over 4 million published scientific articles, we locate around 70,000 articles specialized in MITs or contain relevant data.

With the specialized corpus, we then extract information using NLP methods. Using unsupervised word embedding techniques \cite{ceder-text-19} (such as FastText and word2vec), words are converted to digital tokens that allow streamlined processing. The distribution of word embeddings encodes semantic information. Hence, the similarity (or distance) between words can be quantified, which enables classification, clustering, or identifying novel concepts.
We then apply a pipeline \cite{olivetti-nlp-20} to automatically classify paragraphs in a paper and find those that likely contain pertinent concepts (synthesis recipes, properties, etc.).
As a result, an extended collection of relevant compounds is identified, along with their synthesis recipes (precursors and operations), synthesis conditions (e.g., temperature), and measured properties. When concepts beyond our initial knowledge are discovered, we can define new keywords and repeat the search process iteratively.
With the advancement of large language models (LLMs) and their increasing applications in physical sciences and engineering \cite{llm-science-23, llm-chem-24}, more powerful tools are expected to facilitate this knowledge extraction process.

\subsection{Synthesis Recipe Models}
This literature-extracted compound--property data not only fuels the screening and design of materials, it also allows building ML models for synthesis recipes of materials. This can make a valuable addition to our materials design framework, as the synthesizability of materials determines the realizability of a ``design''.

Our team's initial effort towards this direction \cite{chris-cm-23} explores various ML methods to predict synthesis recipes directly from the chemical formula of precursors and targets. Using the data collected from mining the literature, three supervised ML models are trained: (1) a classification model that predicts the broad category of viable synthesis route; (2) a regression model for synthesis conditions (temperature and reaction time); and (3) a generative model for synthesis conditions.
Using these models, we could predict a viable synthesis recipe given a target material and precursors (e.g., \ch{LiMn_{1.8}Ni_{1.2}O4 <- Li2O3 + Mn2O3 + NiO}): the synthesis route (solid-state, sol-gel, or solution-based reaction), calcination temperature, sintering temperature and time, etc. The models' predictions are validated showing high accuracy in literature-reported synthesis processes, demonstrating practical usefulness in guiding materials synthesis recipe design.

Beyond this preliminary step, the literature-extracted database fosters potential in assessing materials synthesizability, planning synthesis recipes, and facilitating scientific understanding in materials physics and chemistry. As autonomous laboratories are actively developed \cite{autolab-matter-21, autolab-science-24}, synthesis recipe modeling enabled thereby could be a valuable component of these systems.

\section{Virtual Screening}  \label{sec:vscreening}
High-throughput-calculated materials data platforms, such as the Materials Project (MP) \cite{mp-aplm-13} and OQMD \cite{oqmd-jom-13}, contain data for a vast amount of potentially synthesizable materials. 
However, these high-throughput calculations focus on ground-state energies, while transitions driven by temperature and electric fields are not ground-state properties. These databases often contain systematic inaccuracies in electronic and magnetic properties relevant to MIT behavior, which we will discuss in Section \ref{sec:dft}. Despite the limitations, the materials data platforms provide a candidate pool of crystal structures, which we can virtually screen with a predictive model.

\subsection{Interpretable Machine Learning}
From the literature-extracted data, we build an initial database of known materials that display a MIT as a function of temperature, as well as materials that do not display MIT and have only been found in an insulating or a metallic state, but have similar compositions and crystal structures. With this data, we can train a machine learning (ML) predictive model. In this case, we built a classifier predicting whether a material displays an MIT or not. Importantly, such a classifier performs significantly better than a ternary classifier that further differentiates between non-MIT materials and separates them into materials that always display metallic or insulating physical behavior.

In building this ML model, two key issues are (1) data acquisition and (2) data representation. Regarding (1), a structure--property model takes materials structure as input, but many MIT compounds undergo a structural change or distortion upon temperature change, which is sometimes the cause of MIT. Since most materials are first discovered in their high-temperature, high-symmetry phases, we choose only the high-temperature structures that are experimentally validated to train the classifier. This prevents issues such as fusing multi-fidelity data. Nonetheless, leveraging recent advances in multi-fidelity ML \cite{ong-mfgnn-21, chen-mufasa-24}, the restriction could be relaxed, allowing for more flexibility in data sources.

Regarding (2), a representation of materials, i.e., digital formats encoding its composition and crystal structure information, is necessary for ML models. Various materials representation techniques have been developed based on stoichiometry, atomic positions, graph theory, etc., for different tasks. In our virtual screening, we pursue not just predictive power but interpretability, therefore, physically meaningful representations and simple, interpretable ML models are preferred. Physics-based materials descriptors are particularly suitable for this purpose, despite their limitations in design optimization. We adopt the commonly used Magpie descriptors \cite{magpie-npj-16}, which are derived from the composition and atomic properties, and easily obtainable using established toolkits \cite{pymatgen-13, matminer-18}. Represented by descriptors, materials data form a tabular structure where columns are heterogeneous in physical meanings, units, and scales. For tabular data, decision tree-based ensemble models (e.g., gradient boosting) are found to be effective \cite{neurips-tree-22} (also confirmed by our trials of various ML models).

\begin{figure}[ht]
    \centering
    \includegraphics[width=0.5\linewidth]{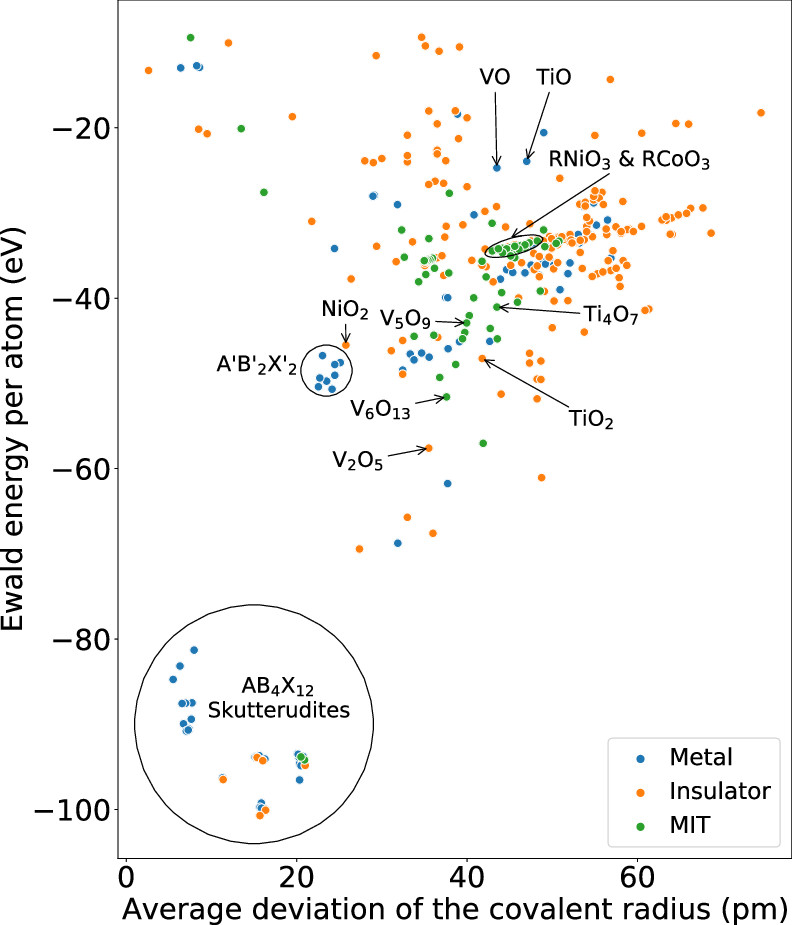}
    \caption{MIT behaviors' relation to Eward energy and ADCR. Reproduced with permission from reference\cite{abg-cm-21}. Copyright 2011 The Authors.}
    \label{fig:key-features}
\end{figure}

With the descriptors as data representation and the tree-based ensemble methods, we have trained a classification model \cite{abg-cm-21} that predicts whether a compound is metal, insulator, or MIT, attaining accuracy comparable to human experts. Moreover, a range of model analysis methods is available to the ensemble models, including the ALE plots \cite{ale-plot-20} that our team developed, and earlier efforts by the ML community, such as the SHAP score \cite{shap-nips-17}. With these methods, we can interpret a trained ML model to assess the importance of descriptors and their effects on the property of interest. It is important in models trained on small datasets that the number of features is not too high to prevent overfitting, to this end, we down-selected to 10 features. Some are related to existing physical models for MIT, including unscreened Hubbard $U$ (related to electron--electron Coulomb interactions) and transition metal distances (related to the kinetic energy of the electrons). The classifier proves its value not just in identifying important features, but also in providing insights to building new physical models. For example, interpreting the classifier identifies that Ewald energy, a measure of ionicity, can be predictive of MIT in binary oxides. Another key feature, the average deviation of the covalent radius (ADCR), which measures the ion size difference in a material, is found to be directly correlated to $T_\mathrm{MIT}$'s within certain materials classes (e.g., rare-earth nickelates \ch{RNiO3}). Figure \ref{fig:key-features} shows the interplay of Eward energy and ADCR in determining the MIT behavior of materials.

With our classification model, we have performed virtual screening of the candidate pool, from which promising candidates of displaying MIT are selected. Upon the results of virtual screening, we build an online database of materials potentially displaying MIT. As these materials' MIT properties are predicted by ML, the fidelity does not yet qualify for experimental verification. Nonetheless, by inspecting the database, we identify materials families promising to host candidates that show desired MIT properties. Within these families, we perform adaptive design optimization to find optimal MIT materials, as described in Section \ref{sec:adaptive}. Meanwhile, we are leveraging physics-based computation for high-fidelity virtual screening.

\subsection{Physics-based Computational Modeling} \label{sec:dft}
For large-scale, high-throughput materials databases, including MP and OQMD, the calculations are often not performed at the required level of physics to understand MITs. In addition, magnetic ordering could significantly affect MIT behaviors. However, MP and OQMD do not consider spin, or assume a simple ferromagnetic configuration, which, for many materials, is not the ground state configuration. For example, in MP, \ch{NdNiO3} is currently listed as a ferromagnetic metal, different from its experimentally observed ground state as an antiferromagnetic insulator.
For these reasons, more accurate models are required to characterize the degrees of freedom, electronic correlations, and resulting MIT behaviors of materials.

In our recent work \cite{alex-24-dftml}, we have used the meta-GGA SCAN exchange-correlation functional to simulate three key materials; this has been found to provide robust results for transition metal oxide ions with partially filled d-orbitals. For the spinel compounds \cite{wang-spinel-prb}, DFT+$U$ has been used, which is faster than meta-GGA functionals. However, the $U$ parameter is not universal and requires appropriate parameterization for a certain materials family. 

Nonetheless, modeling MIT, especially the transition temperature $T_\mathrm{MIT}$, remains difficult: DFT as band theory provides a second-order energy landscape for MIT. DFT+Dynamical Mean Field Theory (DMFT) \cite{alex-commphys-22} is significantly more complex in theory and higher in computational cost. As a result, method development often focuses on well-known materials that can be synthesized with high precision, such as \ch{RNiO3} and \ch{Ca2RuO4}.  Recent works have found that a first-order transition energy landscape can be obtained within DFT+DMFT \cite{alex-pnas-19}. A temperature-induced transition can also be modeled, as DMFT calculations are done at finite electronic temperatures; however, the phonon entropy is still missing, resulting in overestimated $T_\mathrm{MIT}$'s. Models developed as part of this work also allow scientists to separate the role of the crystal lattice, partially filled electron state, and coupling between the two. Despite the low number of MIT materials, we have developed ML models to predict $T_\mathrm{MIT}$'s of possible MIT materials \cite{rf-temperature-24}, and apply them to the materials identified via virtual screening using our classifier.
Separately, MITs may display local variations in $T_\mathrm{MIT}$'s. With the length scale beyond the capability of DFT, we develop phenomenological models with input from ab initio computations, preliminarily demonstrated in synthetic layered materials where local $T_\mathrm{MIT}$'s are controlled \cite{alex-natmater-20, alex-aem-23}.

\section{Adaptive Design Optimization} \label{sec:adaptive}
\subsection{Bayesian Optimization}
ML-based virtual screening allows rapid identification of potential materials families for the target property. Yet, locating the candidate(s) with optimal performance remains a nontrivial effort. Determining the target property with high fidelity requires experiments or physics-based models, particularly DFT.
Investigating the whole materials family experimentally or computationally has a high cost, which is multiplied when more than one property is of interest. Investigating a single crystal structure for a particular stoichiometry with high-fidelity calculations can take days, and identifying the lowest energy crystal structure may take longer, with intense human supervision. Still, one is not guaranteed to identify an MIT material, due to the approximations inherent in DFT, and the lack of temperature dependence in the calculations. Therefore, it is advisable to perform adaptive design: start from small initial data, selectively evaluate new samples, and move towards the optimum.

\begin{figure}[ht]
    \centering
    \includegraphics[width=\textwidth]{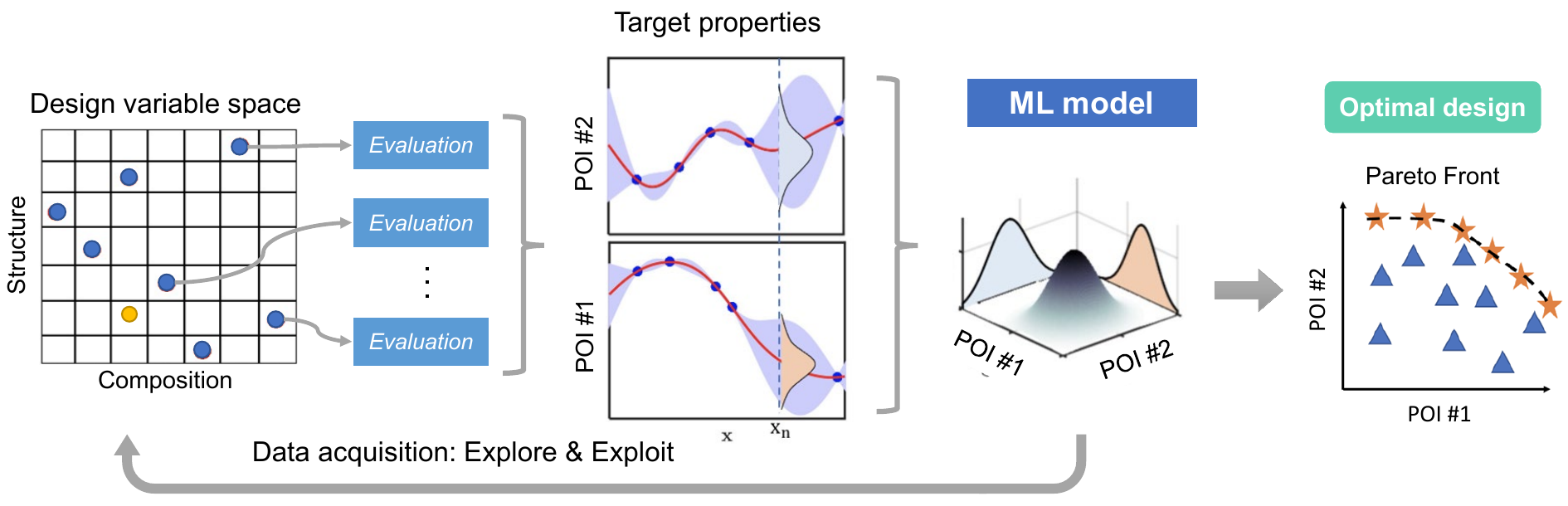}
    \caption{Schematic of BO-based multi-objective adaptive design workflow. A series of initial designs are evaluated with high-fidelity simulations. Therefrom, an ML model is trained, and its predictions guide the acquisition of new designs. The iteration proceeds toward the discovery of the Pareto front.}
    \label{fig:mvbo}
\end{figure}

Bayesian optimization (BO) has been commonly adopted as an efficient and versatile approach to adaptive design. In Figure \ref{fig:mvbo}, we illustrate BO in a general multi-objective design setting. Let $\boldsymbol{x}$ be the design variable that incorporates chemical and structural configuration that can be manipulated; $y_1$, $y_2$ be the properties of interest (POIs) to be optimized. The aim is to seek the Pareto front---where every design is not dominated by others.

Before the BO process, we select a series of initial samples $\{\boldsymbol{x}_1, \dots, \boldsymbol{x}_n\}$ based on prior knowledge, using statistical design of experiments \cite{shang-doe-20}, or from existing data. High-fidelity evaluation methods, such as DFT, are employed to acquire properties for these samples. The subsequent BO process consists of two main components: an uncertainty-aware ML model \cite{zhang-uaml-22} trained on the available data to predict the POIs from the design variables, and an acquisition function that defines a strategy to select a new design to be evaluated. The acquisition function trades off exploration (prioritizes designs with better predicted objective values) against exploitation (prioritizes sampling in the less explored regions of design space). Acquisition functions are usually agnostic about data format or ML model form, while the ML model's compatibility with data limits where BO can be applied. In the following, we present our development of uncertainty-aware ML models to extend BO's applicability.

\subsection{Uncertainty-aware Mixed-variable ML}
Gaussian processes (GPs) \cite{gpml-book} are often the preferred choice of the uncertainty-aware ML models in BO applications, owing to their principled uncertainty quantification (UQ) and modeling flexibility. GPs specify a prior that objective values $\boldsymbol{y}=\{y_1,\dots,y_n\}$ follow a joint Gaussian distribution. As such, GPs are fully characterized by a mean function, which captures a global trend, and a covariance function, which captures the similarity between two observations as a function of the design variables. GPs quantify uncertainty by predicting not just point estimates but distributions for unseen datapoints. This UQ capability is crucial in BO applications, particularly in driving exploration. 

\begin{figure}[ht]
    \centering
    \includegraphics[width=0.8\textwidth]{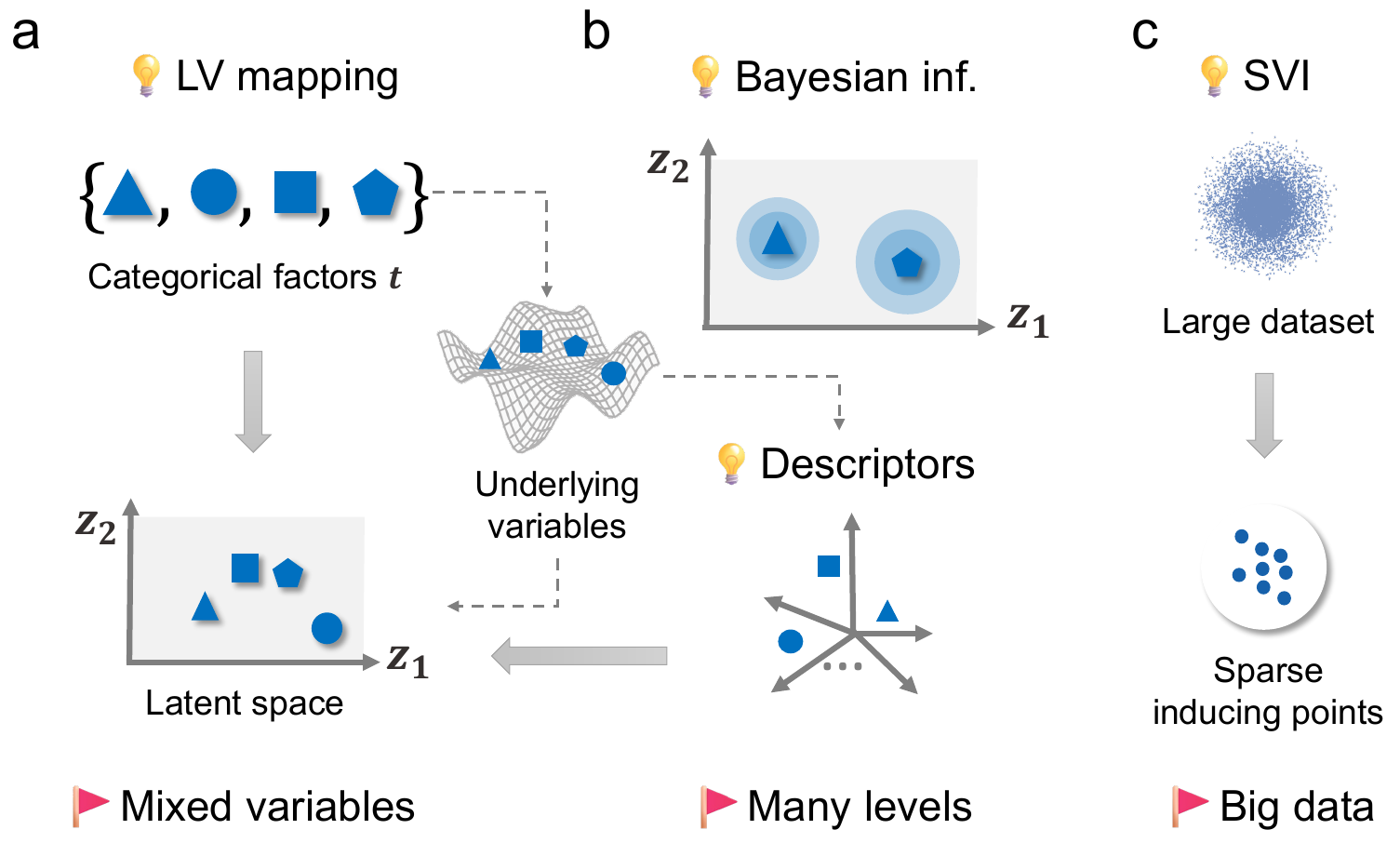}
    \caption{Illustration of the methods extending GP models' capability. (a) The categorical variables are mapped to a latent space that captures the effects of underlying variables, thus enabling mixed-variable modeling \cite{zhang-lvgp-20}. (b) Two approaches are taken to deal with the ``many-level'' challenge: 1) Bayesian inference of latent variable locations \cite{yerramilli-bayesian-2023}; and 2) using descriptors to aid learning latent space configuration \cite{iyer-dalvgp-23}. (c) SVI to overcome the big data challenge \cite{wang-svgp-22}.}
    \label{fig:lvgp-var}
\end{figure}

Despite their advantages, GPs face challenges in modeling mixed-variable design spaces often encountered in materials design scenarios. Standard GP covariance functions based on distances in the design space cannot be directly applied to categorical predictors, as there is no inherent distance metric in the categorical space.
To overcome this challenge and extend the capabilities of the GP models to mixed-variable scenarios, we have developed the latent-variable Gaussian process (LVGP) model \cite{zhang-lvgp-20}. LVGP maps each level of a categorical variable $t$ onto a low-dimensional numerical latent variable (LV) space, as illustrated in Figure \ref{fig:lvgp-var}a. The LV values for each level are estimated from data along with other GP hyperparameters. They quantify the relative distances between the different levels and can be used with standard GP kernel functions. 

Through our BO-based adaptive discovery method, we anticipate uncovering and scrutinizing an expanding array of materials. However, for applications where large data are available or feasible to collect, GPs' scalability becomes an issue. With $N$ datapoints, fitting GP models incurs a computational cost of $O(N^3)$, with memory overhead scaling quadratically. This scaling problem also impacts LVGPs. A viable solution for standard GPs with numerical inputs is to employ stochastic variational inference (SVI) \cite{hensman-svi-13}. SVI approximates the GP model using a smaller set of ``inducing points''. Computational and storage costs are thereby decoupled from the number of datapoints. We have extended SVI to accommodate LVGPs \cite{wang-svgp-22}, addressing challenges in specifying inducing points in the mixed-variable space and optimizing them using modern stochastic gradient methods. This large-scale variant of LVGP, referred to as LVGP-SVI, enables the application of LVGPs to large datasets while preserving the quality of UQ.

In certain mixed-variable materials modeling applications, a challenging scenario arises when categorical variables have many levels. For LVGP to effectively learn the similarity between these levels, it necessitates at least one observation for each category. However, collecting a sufficiently sized training dataset in such scenarios can incur prohibitively high costs. Furthermore, situations may arise where some levels lack any observations altogether.

To address this, we pursued two distinct approaches. Firstly, we introduced a fully Bayesian variant of LVGP \cite{yerramilli-bayesian-2023} to improve learning efficiency during latent variable estimation. Unlike the standard LVGP, this fully Bayesian iteration estimates a posterior distribution for the possible latent variable values, as depicted in Figure \ref{fig:lvgp-var}b. This approach incorporates uncertainty in latent variable estimation into prediction uncertainties, thereby enhancing the quality of UQ crucial for exploration in BO applications. Moreover, it can generate accurate predictions and reliable uncertainty estimates even for levels absent from training data, as demonstrated in empirical studies.

Secondly, we aimed to enhance the model by leveraging domain-specific information, particularly numerical attributes known as descriptors. These descriptors are expected to encapsulate the effects of categorical variable levels on the output. For instance, potential descriptors for a categorical variable representing atom type include atomic weight and covalent radius. We introduced the descriptor-augmented LVGP \cite{iyer-dalvgp-23}, capable of utilizing available descriptors to infer the latent variables, as illustrated in Figure \ref{fig:lvgp-var}b. Empirical results show that even incomplete knowledge of the descriptors can be utilized to reduce the cost of BO for categorical variables with many levels.

These augmentations have equipped the LVGP model with versatile capabilities to adapt to various situations in materials design problems, ranging from small data to large data.

\subsection{Mixed-variable BO for Materials Design}
Empowered by LVGP, we extend the applicability of BO to mixed-variable design spaces. Mixed-variable BO can incorporate various experimental or computational techniques as the ``evaluation'' module. Owing to the wide applicability of LVGP and flexibility of BO, LVGP-enabled mixed-variable BO provides an efficient and generally suitable solution to the design of various materials systems, including dielectric nanocomposite \cite{iyer-msde-20},  solar cell materials \cite{zhang-scirep-20}, and gas-capturing metal--organic frameworks \cite{yc-mof-23}. Built upon mixed-variable BO and DFT calculations, we created an adaptive optimization engine \cite{wang-apr-20} to design materials properties by directly optimizing chemical composition.

\subsection{Optimizing MIT properties}
\begin{figure}[ht]
    \centering
    \includegraphics[width=\textwidth]{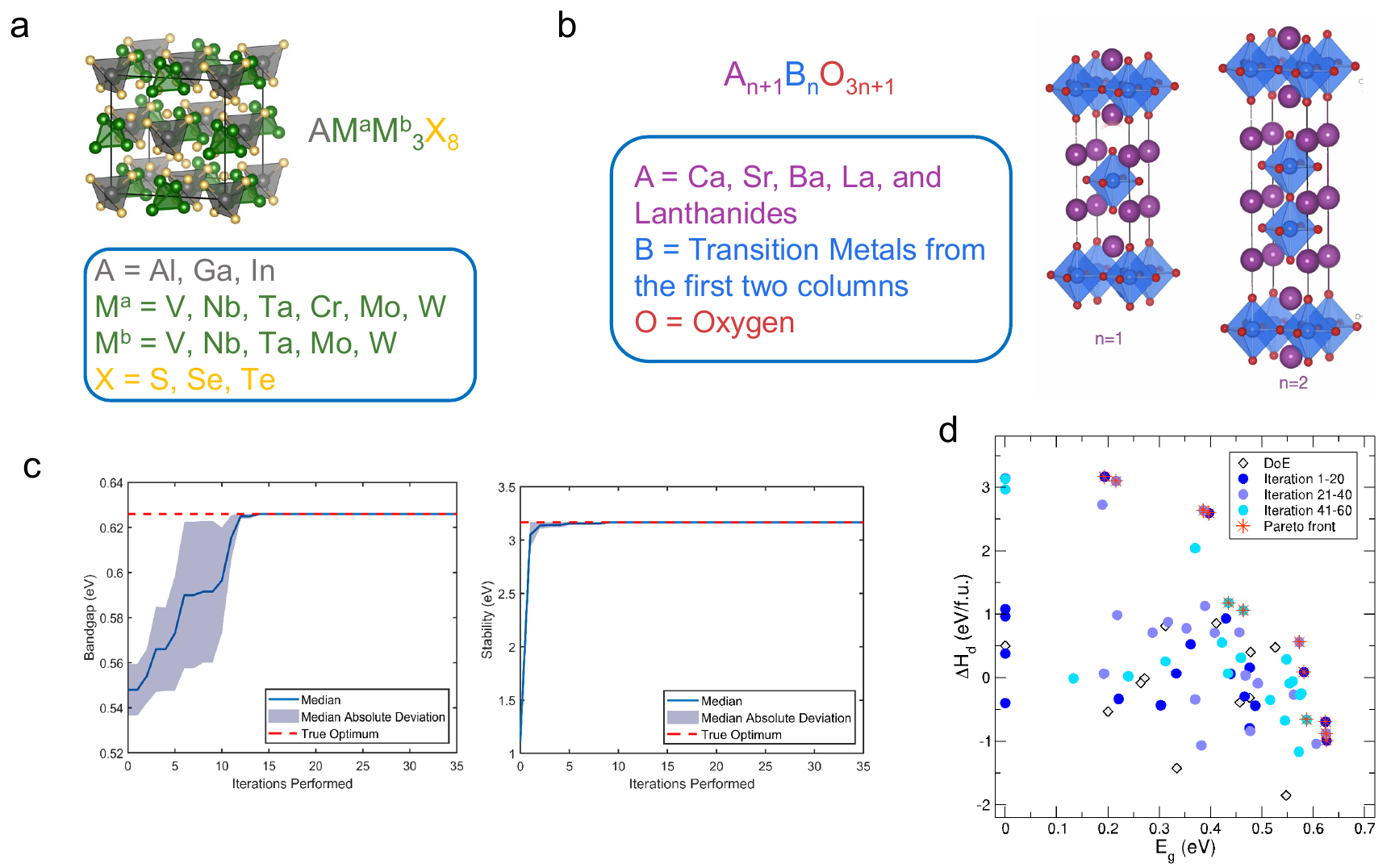}
    \caption{(a,b) Illustration of compositions and structures of the lacunar spinel and Ruddlesden-Popper perovskite families. (c) Single-objective optimization results of bandgap (left) and stability (right). (d) Results of multi-objective optimization. Reproduced with permission from reference \cite{wang-apr-20}. Copyright 2020 The Authors.}
    \label{fig:mater-family}
\end{figure}

In the virtual screening process (Section \ref{sec:vscreening}), we have found materials families of particular interest to MITs, as they contain more materials predicted to be potential MITs than other families.
Examples include (1) lacunar spinels \ch{AM4X8}, with group IIIA metal \ch{A}, transition metals \ch{M4}, and chalcogen \ch{X}; (2) Ruddlesden-Popper perovskites \ch{A_{n+1}B_nO_{3n+1}}, with alkaline earth metals or lanthanides \ch{A} , and transition metal \ch{B}.
Figure \ref{fig:mater-family}a,b show their crystal structures and compositions of interest.

This has reduced the candidates from $10^6$ composition--structures to hundreds of compounds within the candidate families, where searching for the optimal candidates has a manageable cost. We employ BO-based adaptive design (Figure \ref{fig:mvbo}) with DFT as the evaluator to optimize the performance and synthesizability of MITs. We choose the ground state bandgap $E_g$ to indicate MIT performance: a compound with larger $E_g$ generally shows higher resistivity in its insulating state, thus allowing a larger resistivity change ratio upon MIT. Also considered is the stability characterized by decomposition enthalpy $\Delta H_d$, as stable compounds are more likely to be synthesizable and operable in devices. The two properties can be optimized separately using single-objective BO, or concurrently using multi-objective BO.

In Figure \ref{fig:mater-family}c,d, we present an example of optimization within \ch{AM4X8} lacunar spinels \cite{wang-apr-20}. Considering the symmetry requirements of MIT, we reduce the design space to candidates whose \ch{M} sites are occupied by two elements with a 1:3 ratio, i.e., \ch{AM^aM^b3X8}. With 3, 6, 5, and 3 candidate elements on each site, a total of 270 candidates are searched. We represent the candidates by compositions only, using four categorical variables corresponding to element choice at sites, without any hand-crafted features. We demonstrate adaptive design under both the single- and multi-objective settings. In a single-objective setting, starting from 12 initial samples, the global optima candidates of both $E_g$ and $\Delta H_d$ are identified within 15 iterations, exploring less than 10\% of the whole design space (Figure \ref{fig:mater-family}c). Under a multi-objective setting, the first 60 iterations successfully find all candidates on the Pareto front (Figure \ref{fig:mater-family}d). 

Separately, we have also computationally studied the Ruddlesden-Popper \ch{A2BO4}, which may exhibit MIT like that of \ch{Ca2RuO4}, the only known MIT compound in the family.
As a proxy for the insulating state, we simulate the materials in an anti-ferromagnetic state (most likely to have a gap) and a non-magnetic state. As a result, we have identified several possible MIT materials suggested for experimental synthesis. Interestingly, we find the latent variables describing the A-site element related to its atomic radius and ionization state.
Moving forward to materials families with limited prior knowledge, this adaptive design approach could efficiently outline the Pareto front, which contains MIT materials balancing in performance and stability with different emphases, offering options for materials selection according to the application.

\section{Challenges and Outlooks} \label{sec:outlook}
In this Account, we present a framework for systematically accelerating materials design. We have demonstrated the framework's effectiveness in designing MIT materials. The designed materials are predicted to possess promising performances and synthesizability, and they are being experimentally synthesized and characterized by our collaborators for suitability in microelectronic devices.

In the framework, we integrate various AI/ML techniques, including text mining, supervised classification, and mixed variable GP--BO, which address challenges in different phases of 
 materials design process (from conceptual to detailed design), in particular (1) data acquisition from dispersed sources and (2) design optimization in large, disjoint design spaces. In designing synthesizable materials with high MIT performances, the framework has demonstrated progressive knowledge acquisition and effective uncertainty reduction. Moreover, this general framework can be extended and applied to diverse materials design problems.

\begin{figure}[ht]
    \centering
    \includegraphics[width=0.9\textwidth]{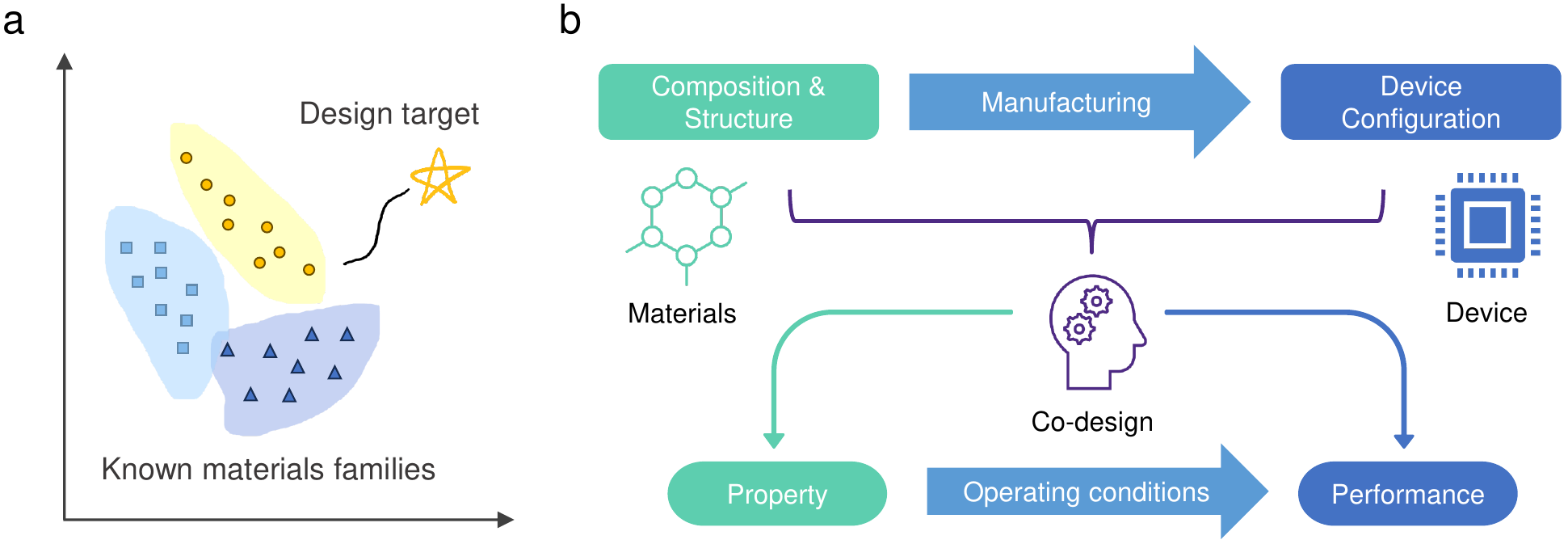}
    \caption{Illustration of the challenges. (a) Data of known materials may not cover the desired target region, limiting the applicability of data-driven models. Reproduced with permission from reference \cite{zhang-scirep-20}. Copyright 2020 The Authors. (b) Device manufacturing requires interaction between material- and device-level objectives and constraints; optimal materials properties do not guarantee desired device performance.}
    \label{fig:outlook}
\end{figure}

Nevertheless, challenges associated with both critical problems mentioned above must be addressed to unlock the full power of data-driven design approaches. Firstly, data quality strongly impacts data-driven studies. Data quality can be assessed from multiple aspects, such as fidelity, diversity, and bias. Both literature mining and virtual screening use the available materials data as is and thus are subject to problems that may present therein, e.g., bias and lack of diversity. As Figure \ref{fig:outlook}a illustrates, if data provided to models do not sufficiently represent the design space, the models will not generalize well and may miss the desired design target. To ensure the reliability of data-driven design, attention should be paid to acquiring and curating high-quality data. There have been efforts toward this end by our team \cite{zhang-etal-23} and others \cite{li-redundancy-23, sparks-notsimple-24}, and we believe more awareness of data quality will make data-driven design workflows more powerful and robust.

Secondly, the presented framework's design optimization focuses on materials properties. In application, the designed MIT materials will be part of microelectronic devices whose holistic performances are the ultimate goal. Ideally, the design of materials and devices should be considered concurrently. To realize ``co-design'', several gaps need to be bridged. One is that the material-level and device-level design problems have interactions. For example, the device fabrication process may require material properties such as mechanical strength and thermal stability. Conversely, the material properties may add constraints to device configuration design. These require information to be propagated between material and device levels in the design framework.

Another gap is the property--performance mismatch: material with optimal properties may not lead to superior device performance. One reason is that the materials' properties are often predicted at the ground state or standard conditions, whereas in real devices the operating conditions vary, e.g., under different environments. Filling this gap requires considering operating conditions in materials property evaluation in the co-design framework.

Finally, going beyond conceptual design, a complete materials development workflow involves validation and deployment. For microelectronic materials, this includes synthesis, characterization, and device fabrication and testing. In these processes, data are generated from both experiments and simulations, manifesting multiple fidelities and modalities. 
We propose the following directions towards integrating experiments and simulations for closed-loop materials development: (1) Fusing multi-fidelity and multimodal data in ML surrogate modeling. (2) Utilizing new data collected from experiments and simulation to ``online'' update the surrogate model. (3) Using model predictive control to facilitate precision synthesis. Beyond guiding adaptive sampling, the model could tune processing conditions in real-time to optimize synthesis quality. With these innovations, we envision a ``digital twin'' for materials design and deployment.

\section*{Acknowledgments}
The information, data, or work presented herein was funded in part by the Advanced Research Projects Agency-Energy (ARPA-E), U.S.~Department of Energy, under award number DE-AR0001209, and the National Science Foundation (NSF), Division of Materials Research, under award number DMR-2324173. H.Z.~acknowledges support from the Ryan Graduate Fellowship. The authors thank Wei ``Wayne'' Chen and Ramin Bostanabad for valuable discussions.

\bibliography{refs}
\end{document}